\PassOptionsToPackage{unicode}{hyperref}
\PassOptionsToPackage{hyphens}{url}
\PassOptionsToPackage{dvipsnames,svgnames,x11names}{xcolor}
\documentclass[
]{article}

\usepackage{amsmath,amssymb}
\usepackage{iftex}
\ifPDFTeX
  \usepackage[T1]{fontenc}
  \usepackage[utf8]{inputenc}
  \usepackage{textcomp} 
\else 
  \usepackage{unicode-math}
  \defaultfontfeatures{Scale=MatchLowercase}
  \defaultfontfeatures[\rmfamily]{Ligatures=TeX,Scale=1}
\fi
\usepackage{lmodern}
\ifPDFTeX\else
\fi
\IfFileExists{upquote.sty}{\usepackage{upquote}}{}
\IfFileExists{microtype.sty}{
  \usepackage[]{microtype}
  \UseMicrotypeSet[protrusion]{basicmath} 
}{}
\makeatletter
\@ifundefined{KOMAClassName}{
  \IfFileExists{parskip.sty}{%
    \usepackage{parskip}
  }{
    \setlength{\parindent}{0pt}
    \setlength{\parskip}{6pt plus 2pt minus 1pt}}
}{
  \KOMAoptions{parskip=half}}
\makeatother
\usepackage{xcolor}
\makeatletter
\ifx\paragraph\undefined\else
  \let\oldparagraph\paragraph
  \renewcommand{\paragraph}{
    \@ifstar
      \xxxParagraphStar
      \xxxParagraphNoStar
  }
  \newcommand{\xxxParagraphStar}[1]{\oldparagraph*{#1}\mbox{}}
  \newcommand{\xxxParagraphNoStar}[1]{\oldparagraph{#1}\mbox{}}
\fi
\ifx\subparagraph\undefined\else
  \let\oldsubparagraph\subparagraph
  \renewcommand{\subparagraph}{
    \@ifstar
      \xxxSubParagraphStar
      \xxxSubParagraphNoStar
  }
  \newcommand{\xxxSubParagraphStar}[1]{\oldsubparagraph*{#1}\mbox{}}
  \newcommand{\xxxSubParagraphNoStar}[1]{\oldsubparagraph{#1}\mbox{}}
\fi
\makeatother

\usepackage{color}
\usepackage{fancyvrb}

\DefineVerbatimEnvironment{Highlighting}{Verbatim}{commandchars=\\\{\}}
\usepackage{framed}
\definecolor{shadecolor}{RGB}{241,243,245}
\newenvironment{Shaded}{\begin{snugshade}}{\end{snugshade}}

\newcommand{\BuiltInTok}[1]{\textcolor[rgb]{0.00,0.23,0.31}{#1}}

\newcommand{\ConstantTok}[1]{\textcolor[rgb]{0.56,0.35,0.01}{#1}}

\newcommand{\FloatTok}[1]{\textcolor[rgb]{0.68,0.00,0.00}{#1}}
\newcommand{\FunctionTok}[1]{\textcolor[rgb]{0.28,0.35,0.67}{#1}}
\newcommand{\ImportTok}[1]{\textcolor[rgb]{0.00,0.46,0.62}{#1}}

\newcommand{\KeywordTok}[1]{\textcolor[rgb]{0.00,0.23,0.31}{\textbf{#1}}}
\newcommand{\NormalTok}[1]{\textcolor[rgb]{0.00,0.23,0.31}{#1}}
\newcommand{\OperatorTok}[1]{\textcolor[rgb]{0.37,0.37,0.37}{#1}}

\newcommand{\PreprocessorTok}[1]{\textcolor[rgb]{0.68,0.00,0.00}{#1}}

\newcommand{\StringTok}[1]{\textcolor[rgb]{0.13,0.47,0.30}{#1}}

\usepackage{longtable,booktabs,array}
\usepackage{calc} 
\usepackage{etoolbox}
\makeatletter
\patchcmd\longtable{\par}{\if@noskipsec\mbox{}\fi\par}{}{}
\makeatother
\IfFileExists{footnotehyper.sty}{\usepackage{footnotehyper}}{\usepackage{footnote}}
\makesavenoteenv{longtable}
\usepackage{graphicx}
\makeatletter
\newsavebox\pandoc@box
\newcommand*\pandocbounded[1]{
  \sbox\pandoc@box{#1}%
  \Gscale@div\@tempa{\textheight}{\dimexpr\ht\pandoc@box+\dp\pandoc@box\relax}%
  \Gscale@div\@tempb{\linewidth}{\wd\pandoc@box}%
  \ifdim\@tempb\p@<\@tempa\p@\let\@tempa\@tempb\fi
  \ifdim\@tempa\p@<\p@\scalebox{\@tempa}{\usebox\pandoc@box}%
  \else\usebox{\pandoc@box}%
  \fi%
}
\def\fps@figure{htbp}
\makeatother
\NewDocumentCommand\citeproctext{}{}

\makeatletter
 \let\@cite@ofmt\@firstofone
 \def\@biblabel#1{}
 \def\@cite#1#2{{#1\if@tempswa , #2\fi}}
\makeatother
\newlength{\cslhangindent}
\newlength{\csllabelwidth}
\newenvironment{CSLReferences}[2] 
 {\begin{list}{}{%
  \setlength{\itemindent}{0pt}
  \setlength{\leftmargin}{0pt}
  \setlength{\parsep}{0pt}
  \ifodd #1
   \setlength{\leftmargin}{\cslhangindent}
   \setlength{\itemindent}{-1\cslhangindent}
  \fi
  \setlength{\itemsep}{#2\baselineskip}}}
 {\end{list}}
\usepackage{calc}

\usepackage{arxiv}
\usepackage{orcidlink}
\usepackage{amsmath}
\usepackage[T1]{fontenc}
\makeatletter
\@ifpackageloaded{caption}{}{\usepackage{caption}}
\AtBeginDocument{%
\ifdefined\contentsname
  \renewcommand*\contentsname{Table of contents}
\else
  \newcommand\contentsname{Table of contents}
\fi
\ifdefined\listfigurename
  \renewcommand*\listfigurename{List of Figures}
\else
  \newcommand\listfigurename{List of Figures}
\fi
\ifdefined\listtablename
  \renewcommand*\listtablename{List of Tables}
\else
  \newcommand\listtablename{List of Tables}
\fi
\ifdefined\figurename
  \renewcommand*\figurename{Figure}
\else
  \newcommand\figurename{Figure}
\fi
\ifdefined\tablename
  \renewcommand*\tablename{Table}
\else
  \newcommand\tablename{Table}
\fi
}
\@ifpackageloaded{float}{}{\usepackage{float}}
\floatstyle{ruled}
\@ifundefined{c@chapter}{\newfloat{codelisting}{h}{lop}}{\newfloat{codelisting}{h}{lop}[chapter]}
\floatname{codelisting}{Listing}

\makeatother
\makeatletter
\@ifpackageloaded{caption}{}{\usepackage{caption}}
\@ifpackageloaded{subcaption}{}{\usepackage{subcaption}}
\makeatother

\usepackage{bookmark}

\IfFileExists{xurl.sty}{\usepackage{xurl}}{} 
\hypersetup{
  pdftitle={IdentityByDescentDispersal.jl: Inferring dispersal rates with identity-by-descent blocks},
  pdfkeywords={population genetics, identity-by-descent, Julia},
  colorlinks=true,
  linkcolor={blue},
  filecolor={Maroon},
  citecolor={Blue},
  urlcolor={Blue},
  pdfcreator={LaTeX via pandoc}}

\newcommand{\runninghead}{A Preprint }
\renewcommand{\runninghead}{IdentityByDescentDispersal.jl }
\title{IdentityByDescentDispersal.jl: Inferring dispersal rates with
identity-by-descent blocks}
\def\asep{\\\\\\ } 
\author{\textbf{Francisco Campuzano
Jiménez}~\orcidlink{0000-0001-8285-9318}\\Evolutionary Ecology Group,
Department of Biology\\University of Antwerp,
Belgium\\\\\href{mailto:Curro.CampuzanoJimenez@uantwerpen.be}{Curro.CampuzanoJimenez@uantwerpen.be}\asep\textbf{Arthur
Zwaenepoel}~\orcidlink{0000-0003-1085-2912}\\Evolutionary Ecology Group,
Department of Biology\\University of Antwerp,
Belgium\\\\\href{mailto:Arthur.Zwaenepoel@uantwerpen.be}{Arthur.Zwaenepoel@uantwerpen.be}\asep\textbf{Els
Lea R De Keyzer}~\orcidlink{0000-0003-0924-0118}\\Evolutionary Ecology
Group, Department of Biology\\University of Antwerp,
Belgium\\\\\href{mailto:Els.DeKeyzer@uantwerpen.be}{Els.DeKeyzer@uantwerpen.be}\asep\textbf{Hannes
Svardal}~\orcidlink{0000-0001-7866-7313}\\Evolutionary Ecology Group,
Department of Biology\\University of Antwerp, Belgium\\\\\\Naturalis
Biodiversity Center, Leiden,
Netherlands\\\\\href{mailto:Hannes.Svardal@uantwerpen.be}{Hannes.Svardal@uantwerpen.be}}
\date{}
\begin{document}
\maketitle
{\bfseries \emph Keywords}
\def\sep{\textbullet\ }
population genetics \sep identity-by-descent \sep
Julia

\section{Summary}\label{summary}

The population density and per-generation dispersal rate of a population
are central parameters in the study of evolution and ecology. The
dispersal rate is particularly relevant for conservation management of
fragmented or invasive species (Driscoll et al. 2014). There is a
growing interest in developing statistical methods that exploit the
increasingly available genetic data to estimate the effective population
density and effective dispersal rate (Rousset 1997; Ringbauer, Coop, and
Barton 2017; Chris C. R. Smith et al. 2023; Chris C. R. Smith and Kern
2023).

The distribution of recent coalescent events between individuals in
space can be used to estimate such quantities through the distribution
of identity-by-descent (IBD) blocks (Ringbauer, Coop, and Barton 2017).
An IBD block is defined as a segment of DNA that has been inherited by a
pair of individuals from a common ancestor without being broken by
recombination. Here we present \texttt{IdentityByDescentDispersal.jl}, a
Julia package for estimating effective population densities and
dispersal rates from observed spatial patterns of IBD shared blocks.

\section{Statement of need}\label{sec-statement-of-need}

Ringbauer, Coop, and Barton (2017) proposed an inference scheme for the
estimation of effective population density and effective dispersal rate
from shared IBD blocks. Despite their promising results, there is to
this date no general-purpose software implementation of their method.

In order to make the inference approach available to the broader
audience of evolutionary biologists and conservation scientists, we
present \texttt{IdentityByDescentDispersal.jl}, a Julia (Bezanson et al.
2017) package with an efficient and easy-to-use implementation of the
method. The package implements the core equations proposed by Ringbauer,
Coop, and Barton (2017) and can be used to perform composite
likelihood-based inference using either maximum-likelihood estimation
(MLE) or Bayesian inference.

The method of Ringbauer, Coop, and Barton (2017) was limited to a family
of functions for the change in effective population density over time of
the form \(D_e(t) = Dt^{-\beta}\), for which the theory was analytically
tractable. In addition, in the paper describing the original approach,
the authors used gradient-free optimization to calculate maximum
likelihood estimates (MLEs). Our implementation makes two major software
contributions. First, we admit composite likelihood calculations for
arbitrary functions \(D_e(t)\) by evaluating the relevant integrals
numerically through Gaussian quadrature rules (Johnson 2013). This
significantly enlarges the space of biologically relevant models that
can be fitted. Second, our implementation takes advantage of the
powerful Julia ecosystem and the work of Geoga et al. (2022) to provide
a version of the composite likelihood that is fully compatible with
automatic differentiation (AD), including AD with respect to \(\beta\).
By having a fully AD-compatible composite likelihood,
\texttt{IdentityByDescentDispersal.jl} can be used together with
standard gradient-based optimization and sampling methods available in
the Julia ecosystem, which are typically more efficient than
gradient-free methods.

Lastly, our package comes with a template to simulate synthetic datasets
and a pipeline for end-to-end analysis from VCF files to final
estimates. We believe it will encourage a broader audience to adopt the
inference scheme proposed by Ringbauer, Coop, and Barton (2017),
motivate further developments and expand its applications.

\section{Overview}\label{sec-overview}

\texttt{IdentityByDescentDispersal.jl} contains two main sets of
functions. The first set has the prefix \texttt{expected\_ibd\_blocks}
and allows users to calculate the expected density of IBD blocks per
pair of individuals and per unit of block length for various demographic
models by solving

\begin{equation}\phantomsection\label{eq-1}{
\mathbb{E}[N_L | r, \theta] = \int_0^\infty G4t^2 \exp(-2Lt) \cdot \Phi(t | r, \theta) \,dt
}\end{equation} where \(G\) is the length of the genome (in Morgan),
\(t\) is time (generations in the past), \(L\) is the length of the
block (Morgan) and \(r\) is the geographical distance in the present (at
time \(t=0\)) between the two individuals. \(\Phi(t| r, \theta)\) is the
instantaneous coalescence rate at time \(t\) of two homologous loci that
are initially \(r\) units apart under the demographic model with
parameters \(\theta\). A slightly more complicated expression that
accounts for chromosomal edges and diploidy is the default in
\texttt{IdentityByDescentDispersal.jl}.

The second set of functions has the prefix
\texttt{composite\_loglikelihood} and allows users to directly compute
the composite likelihood of the data by assuming the observed number of
IBD blocks whose lengths fall in a small bin \([L, L+\Delta L]\) and are
shared by a pair of individuals \(r\) units apart follows a Poisson
distribution with mean
\(\lambda = \mathbb{E}[N_L | r, \theta] \Delta L\).

\texttt{IdentityByDescentDispersal.jl} allows for three different parameterizations of the effective population density function:
a constant density, a power-density, and a user-defined density (see \autoref{tbl-tab1}).

\begin{longtable}[]{@{}
  >{\raggedright\arraybackslash}p{(\linewidth - 6\tabcolsep) * \real{0.2400}}
  >{\raggedright\arraybackslash}p{(\linewidth - 6\tabcolsep) * \real{0.2667}}
  >{\raggedright\arraybackslash}p{(\linewidth - 6\tabcolsep) * \real{0.3333}}
  >{\raggedright\arraybackslash}p{(\linewidth - 6\tabcolsep) * \real{0.1600}}@{}}
\caption{\texttt{IdentityByDescentDispersal.jl} functions support three
different parameterizations that are indicated by their respective
suffixes. \label{tbl-tab1}}\tabularnewline
\toprule\noalign{}
\begin{minipage}[b]{\linewidth}\raggedright
Function suffix
\end{minipage} & \begin{minipage}[b]{\linewidth}\raggedright
\(D_e(t)\) formula
\end{minipage} & \begin{minipage}[b]{\linewidth}\raggedright
Parameters
\end{minipage} & \begin{minipage}[b]{\linewidth}\raggedright
Solver
\end{minipage} \\
\midrule\noalign{}
\endfirsthead
\toprule\noalign{}
\begin{minipage}[b]{\linewidth}\raggedright
Function suffix
\end{minipage} & \begin{minipage}[b]{\linewidth}\raggedright
\(D_e(t)\) formula
\end{minipage} & \begin{minipage}[b]{\linewidth}\raggedright
Parameters
\end{minipage} & \begin{minipage}[b]{\linewidth}\raggedright
Solver
\end{minipage} \\
\midrule\noalign{}
\endhead
\bottomrule\noalign{}
\endlastfoot
\texttt{constant\_density} & \(D_e(t)=D\) & \(D,\ \sigma\) &
Analytically \\
\texttt{power\_density} & \(D_e(t)=Dt^{-\beta}\) &
\(D,\ \beta,\ \sigma\) & Analytically \\
\texttt{custom} & User-defined & User-defined and \(\sigma\) &
Numerically \\
\end{longtable}

The Julia package is accompanied by two additional resources. First, we
provide a simulation template in SLiM for forward-in-time population
genetics simulation in a continuous space with tree-sequence recording
(Haller and Messer 2023; Haller et al. 2019). This template can be used
to assess model assumptions, guide empirical analysis, and perform
simulation-based calibration. Assessing the performance of the method
with synthetic datasets is a crucial step, as it is known that errors in
the detection of IBD blocks are common (S. R. Browning and Browning
2012) and that inferences based on composite likelihood tend to be
overconfident, underestimating posterior uncertainty and yielding too
narrow confidence intervals.

Second, we have also implemented a bioinformatics pipeline that carries
out a complete analysis from detecting IBD blocks to finding the MLE of
the effective population density and the effective dispersal rate. It is
shared as a Snakemake pipeline, a popular bioinformatics workflow
management tool (Mölder et al. 2021). It takes as input a set of phased
VCF files, their corresponding genetic maps and a CSV file containing
pairwise geographical distances between individuals. The pipeline
detects IBD blocks using HapIBD (Zhou, Browning, and Browning 2020),
post-processes them with Refined IBD (B. L. Browning and Browning 2013)
and produces a CSV file compatible with subsequent analysis with
\texttt{IdentityByDescentDispersal.jl} via the
\texttt{preprocess\_dataset} function.

Both the SLiM simulation template and the Snakemake pipeline can be
found in the GitHub repository at
\url{https://github.com/currocam/IdentityByDescentDispersal.jl}.

\section{Example}\label{sec-example}

In this section, we demonstrate how
\texttt{IdentityByDescentDispersal.jl} can be used together with the
popular \texttt{Turing.jl} framework (Ge, Xu, and Ghahramani 2018) using
a dataset we simulate in the documentation. We analyze error-free IBD
blocks shared by 100 diploid individuals from a constant-density
population with parameters \(D_{\text{true}}\approx 250\)
diploids/km\textsuperscript{2} and \(\sigma_{\text{true}}\approx 0.071\)
km/generation.

\texttt{IdentityByDescentDispersal.jl} has extensive documentation that
covers the underlying theory behind the method, how to effectively
simulate synthetic datasets, various demographic models, and inference
algorithms. We refer the reader to the documentation for more details,
which can be found at
\href{https://currocam.github.io/IdentityByDescentDispersal.jl/dev/}{https://currocam.github.io/IdentityByDescentDispersal.jl/}.

\begin{figure}[h]

\centering{

\includegraphics[width=0.85\linewidth,height=\textheight,keepaspectratio]{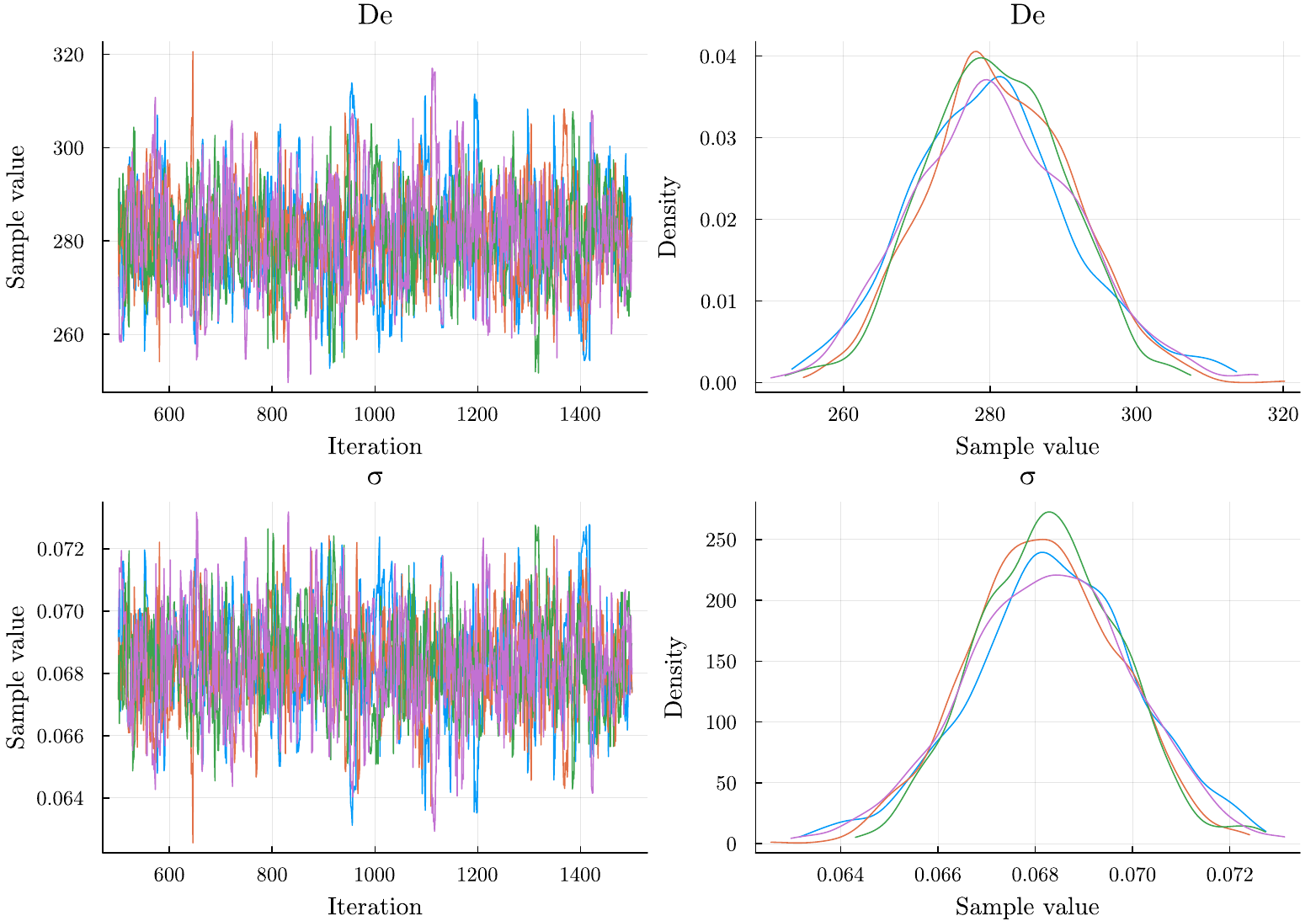}

}

\caption{\label{fig-example}Estimated pseudo-posterior obtained by doing
inference with the composite likelihood. The pseudo-posterior is forced
to concentrate near the true values
(\(\mathbb E[D | \text{data}]\approx 281\) and
\(\mathbb E[\sigma | \text{data}]\approx 0.068\), respectively).}

\end{figure}

Thanks to \texttt{Turing.jl}, we can perform Bayesian inference with a
wide range of popular Monte Carlo algorithms. \autoref{fig-example}
shows the estimated pseudo-posterior obtained through doing inference
with the composite likelihood.

\autoref{fig-example} was generated by the following snippet of Julia
code, which reads the processed data CSV from the provided Snakemake
pipeline.

\begin{Shaded}
\begin{Highlighting}[]
\ImportTok{using} \BuiltInTok{CSV}\NormalTok{, }\BuiltInTok{DataFrames}\NormalTok{, }\BuiltInTok{Turing}\NormalTok{, }\BuiltInTok{StatsPlots}\NormalTok{, }\BuiltInTok{IdentityByDescentDispersal}
\NormalTok{df }\OperatorTok{=}\NormalTok{ CSV.}\FunctionTok{read}\NormalTok{(}\StringTok{"ibd\_dispersal\_data.csv"}\NormalTok{, DataFrame)}
\NormalTok{contig\_lengths }\OperatorTok{=}\NormalTok{ [}\FloatTok{1.0}\NormalTok{]}
\PreprocessorTok{@model} \KeywordTok{function} \FunctionTok{constant\_density}\NormalTok{(df, contig\_lengths)}
\NormalTok{    De }\OperatorTok{\textasciitilde{}} \FunctionTok{Truncated}\NormalTok{(}\FunctionTok{Normal}\NormalTok{(}\FloatTok{1000}\NormalTok{, }\FloatTok{100}\NormalTok{), }\FloatTok{0}\NormalTok{, }\ConstantTok{Inf}\NormalTok{)}
\NormalTok{    sigma }\OperatorTok{\textasciitilde{}} \FunctionTok{InverseGamma}\NormalTok{(}\FloatTok{1}\NormalTok{, }\FloatTok{1}\NormalTok{)}
\NormalTok{    Turing.}\PreprocessorTok{@addlogprob}\NormalTok{! }\FunctionTok{composite\_loglikelihood\_constant\_density}\NormalTok{(}
\NormalTok{      De, sigma, df, contig\_lengths}
\NormalTok{    )}
\KeywordTok{end}
\NormalTok{m }\OperatorTok{=} \FunctionTok{constant\_density}\NormalTok{(df, contig\_lengths)}
\NormalTok{chains }\OperatorTok{=} \FunctionTok{sample}\NormalTok{(m, }\FunctionTok{NUTS}\NormalTok{(), }\FunctionTok{MCMCThreads}\NormalTok{(), }\FloatTok{1000}\NormalTok{, }\FloatTok{4}\NormalTok{)}
\FunctionTok{plot}\NormalTok{(chains)}
\end{Highlighting}
\end{Shaded}

We can also easily compute the MLEs of the same demographic model,

\begin{Shaded}
\begin{Highlighting}[]
\NormalTok{mle\_estimate }\OperatorTok{=} \FunctionTok{maximum\_likelihood}\NormalTok{(}
\NormalTok{  m; lb}\OperatorTok{=}\NormalTok{[}\FloatTok{0.0}\NormalTok{, }\FloatTok{0.0}\NormalTok{], ub}\OperatorTok{=}\NormalTok{[}\FloatTok{1e8}\NormalTok{, }\FloatTok{1e8}\NormalTok{]}
\NormalTok{)}
\FunctionTok{coeftable}\NormalTok{(mle\_estimate)}
\end{Highlighting}
\end{Shaded}

which estimates \(D_{\text{MLE}}\approx 282\) diploids/km² (95\% CI:
260--303) and \(\sigma_{\text{MLE}}\approx 0.068\) km/generation (95\%
CI: 0.065--0.071). The 95\% confidence interval is computed from the
Fisher information matrix.

\section*{Availability}\label{availability}
\addcontentsline{toc}{section}{Availability}

\texttt{IdentityByDescentDispersal.jl} is a registered Julia package
available through the official General registry. Its source code is
hosted on GitHub at
\url{https://github.com/currocam/IdentityByDescentDispersal.jl}.

\section*{Acknowledgements}\label{acknowledgements}
\addcontentsline{toc}{section}{Acknowledgements}

We acknowledge financial support from the Research Foundation - Flanders
(FWO). This work was supported by FWO-G0A9B24N (F.C.J, H.S),
FWO-1272625N (A.Z., H.S) and FWO-12A8423N (E.L.R.D.K., H.S).

\section*{References}\label{references}
\addcontentsline{toc}{section}{References}

\phantomsection\label{refs}


\begin{CSLReferences}{1}{0}
\bibitem[\citeproctext]{ref-bezanson_julia_2017}
Bezanson, Jeff, Alan Edelman, Stefan Karpinski, and Viral B. Shah. 2017.
{``Julia: {A} {Fresh} {Approach} to {Numerical} {Computing}.''}
\emph{SIAM Review} 59 (1): 65--98.
\url{https://doi.org/10.1137/141000671}.

\bibitem[\citeproctext]{ref-browning_improving_2013}
Browning, Brian L., and Sharon R. Browning. 2013. {``Improving the
Accuracy and Efficiency of Identity-by-Descent Detection in Population
Data.''} \emph{Genetics} 194 (2): 459--71.
\url{https://doi.org/10.1534/genetics.113.150029}.

\bibitem[\citeproctext]{ref-browning_identity_2012}
Browning, Sharon R., and Brian L. Browning. 2012. {``Identity by Descent
Between Distant Relatives: Detection and Applications.''} \emph{Annual
Review of Genetics} 46: 617--33.
\url{https://doi.org/10.1146/annurev-genet-110711-155534}.

\bibitem[\citeproctext]{ref-driscoll_trajectory_2014}
Driscoll, Don A., Sam C. Banks, Philip S. Barton, Karen Ikin, Pia
Lentini, David B. Lindenmayer, Annabel L. Smith, et al. 2014. {``The
{Trajectory} of {Dispersal} {Research} in {Conservation} {Biology}.
{Systematic} {Review}.''} \emph{PLOS ONE} 9 (4): e95053.
\url{https://doi.org/10.1371/journal.pone.0095053}.

\bibitem[\citeproctext]{ref-ge_turing_2018}
Ge, Hong, Kai Xu, and Zoubin Ghahramani. 2018. {``Turing: {A} {Language}
for {Flexible} {Probabilistic} {Inference}.''} In \emph{Proceedings of
the {Twenty}-{First} {International} {Conference} on {Artificial}
{Intelligence} and {Statistics}}, 1682--90. PMLR.
\url{https://proceedings.mlr.press/v84/ge18b.html}.

\bibitem[\citeproctext]{ref-geoga_fitting_2022}
Geoga, Christopher J., Oana Marin, Michel Schanen, and Michael L. Stein.
2022. {``Fitting {Matérn} {Smoothness} {Parameters} {Using} {Automatic}
{Differentiation}.''} arXiv.
\url{https://doi.org/10.48550/ARXIV.2201.00090}.

\bibitem[\citeproctext]{ref-haller_tree-sequence_2019}
Haller, Benjamin C., Jared Galloway, Jerome Kelleher, Philipp W. Messer,
and Peter L. Ralph. 2019. {``Tree-Sequence Recording in {SLiM} Opens New
Horizons for Forward-Time Simulation of Whole Genomes.''}
\emph{Molecular Ecology Resources} 19 (2): 552--66.
\url{https://doi.org/10.1111/1755-0998.12968}.

\bibitem[\citeproctext]{ref-haller_slim_2023}
Haller, Benjamin C., and Philipp W. Messer. 2023. {``{SLiM} 4:
{Multispecies} {Eco}-{Evolutionary} {Modeling}.''} \emph{The American
Naturalist} 201 (5): E127--39. \url{https://doi.org/10.1086/723601}.

\bibitem[\citeproctext]{ref-quadgk}
Johnson, Steven G. 2013. {``{QuadGK.jl}: {G}auss--{K}ronrod Integration
in {J}ulia.''} \url{https://github.com/JuliaMath/QuadGK.jl}.

\bibitem[\citeproctext]{ref-molder_sustainable_2021}
Mölder, Felix, Kim Philipp Jablonski, Brice Letcher, Michael B. Hall,
Christopher H. Tomkins-Tinch, Vanessa Sochat, Jan Forster, et al. 2021.
{``Sustainable Data Analysis with {Snakemake}.''} \emph{F1000Research}
10: 33. \url{https://doi.org/10.12688/f1000research.29032.2}.

\bibitem[\citeproctext]{ref-ringbauer_inferring_2017}
Ringbauer, Harald, Graham Coop, and Nicholas H. Barton. 2017.
{``Inferring {Recent} {Demography} from {Isolation} by {Distance} of
{Long} {Shared} {Sequence} {Blocks}.''} \emph{Genetics} 205 (3):
1335--51. \url{https://doi.org/10.1534/genetics.116.196220}.

\bibitem[\citeproctext]{ref-rousset_genetic_1997}
Rousset, François. 1997. {``Genetic {Differentiation} and {Estimation}
of {Gene} {Flow} from {F}-{Statistics} {Under} {Isolation} by
{Distance}.''} \emph{Genetics} 145 (4): 1219--28.
\url{https://doi.org/10.1093/genetics/145.4.1219}.

\bibitem[\citeproctext]{ref-smith_dispersal_2023}
Smith, Chris C R, Silas Tittes, Peter L Ralph, and Andrew D Kern. 2023.
{``Dispersal Inference from Population Genetic Variation Using a
Convolutional Neural Network.''} \emph{Genetics} 224 (2): iyad068.
\url{https://doi.org/10.1093/genetics/iyad068}.

\bibitem[\citeproctext]{ref-smith_dispersenn2_2023}
Smith, Chris C. R., and Andrew D. Kern. 2023. {``{disperseNN2}: A Neural
Network for Estimating Dispersal Distance from Georeferenced
Polymorphism Data.''} \emph{BMC Bioinformatics} 24 (1): 385.
\url{https://doi.org/10.1186/s12859-023-05522-7}.

\bibitem[\citeproctext]{ref-zhou_fast_2020}
Zhou, Ying, Sharon R. Browning, and Brian L. Browning. 2020. {``A {Fast}
and {Simple} {Method} for {Detecting} {Identity}-by-{Descent} {Segments}
in {Large}-{Scale} {Data}.''} \emph{American Journal of Human Genetics}
106 (4): 426--37. \url{https://doi.org/10.1016/j.ajhg.2020.02.010}.

\end{CSLReferences}
\end{document}